\documentstyle[lprocldpf,epsf]{article}

\def\pmb#1{\setbox0=\hbox{#1}%
  \kern-.025em\copy0\kern-\wd0
  \kern.05em\copy0\kern-\wd0
  \kern-.025em\raise.0433em\box0 }

\def\nang{\langle n\rangle}

\def\dxy{d_{x^2-y^2}}

\def\lb{\pmb{$\ell$}}
\def\pp{(\pi,\pi)}

\def\txN{{\textstyle{1\over N}}}
\def\txN2{\textstyle{1\over N^2}}

\def\im{{\rm Im\,}}

\def\cuo2{CuO$_2$}

\def\2kf{2\,k_F}
\def\4kf{4\,k_F}

\def\q{{\bf q}}
\def\p{{\bf p}}

\def\om{\omega}

\def\Journal#1#2#3#4{{#1} {\bf #2}, #3 (#4)}

\def\PRL{\em Phys. Rev. Lett.}
\def\PRB{{\em Phys. Rev.} B}

\def\p{{\bf p}}
\def\q{{\bf q}}
\def\om{\omega}

\begin{document}

\title{PAIRING MECHANISM IN THE TWO--DIMENSIONAL HUBBARD MODEL}

\author{D.J. SCALAPINO}

\address{Department of Physics, University of California,
Santa Barbara,\\ CA 93106--9530, USA}


\vspace{0.5cm}
\maketitle\abstracts{Here we discuss Quantum Monte Carlo results 
for the magnetic susceptibility,
single--particle spectral weight and the irreducible particle--particle 
interaction vertex of the two--dimensional Hubbard model.
In the doped system, as the temperature is lowered below $J=4\,t^2/U$,
short--range antiferromagnetic correlations develop.
These lead to a narrow low--energy quasiparticle band with a large Fermi
surface and a particle--particle vertex which increases at large momentum 
transfer, which favor $\dxy$--pairing.
}


A variety of experiments on the high $T_c$ cuprates can be interpreted
in terms of a gap with dominant $\dxy$ symmetry~\cite{Sch}.
However, the implications of this with respect to the nature of the
underlying pairing mechanism remains an open question.
Here we review some results which have been obtained for the
two--dimensional Hubbard model, which has been found to exhibit 
$\dxy$--like pairing fluctuations.
Our aim is to examine in this particular case the structure of the
quasiparticle spectrum and the interaction in order to gain insight
into the mechanism which leads to $\dxy$ pairing correlations in this
model.


The two-dimensional Hubbard Hamiltonian provides an approximate 
model of a CuO$_2$ layer:
\begin{equation}
H=-t\sum_{\langle ij\rangle ,s} \left( c^\dagger_{is}c_{js} +
c^\dagger_{js}c_{is} \right) + U \sum_i n_{i\uparrow}n_{i\downarrow}.
\label{eq:ham}
\end{equation}
Here $c^\dagger_{is}$ creates an electron of spin $s$ on site $i$, 
and $n_{is} =
c^\dagger_{is}c_{is}$ is the occupation number for spin $s$ on site $i$.
The one--electron transfer between near--neighbor sites is $t$, and $U$ is
an onsite Coulomb energy.
The bare energy scale is set by the bandwidth $8\,t$ and the effective
Coulomb interaction $U$, which are both of order electron volts.
Near half--filling, electrons on neighboring sites tend to align
antiferromagnetically so as to lower their energy by the exchange
interaction $J=4\,t^2/U$.
This interaction is of order a tenth of an electron volt and, as we will
see, sets the energy, or temperature scale, below which
antiferromagnetic (AF) correlations, the low--energy structure in the
single--particle spectral weight, and the pairing interaction develop.

While Monte Carlo~\cite{Dag} and Lanczos~\cite{Par} calculations for a 
$4\times4$ lattice find that two holes
added to the half--filled Hubbard ground state form a $\dxy$ bound state,
and density matrix renormalization group calculations~\cite{Noack} find that
$\dxy$--like pairs are formed on two--leg Hubbard ladders, it is not known
what happens for the two-dimensional Hubbard model.
It is possible that on an energy scale of order $J/10$, a $\dxy$
superconducting state forms.
However, this may well require modifications of the model, such as an
additional near--neighbor $\Delta J {\bf S}_i\cdot{\bf S}_j$ term or
possibly a next--near--neighbor hopping $t'$.
Nevertheless, it is known that as the temperature is reduced below $J$,
$\dxy$ pairing correlations develop in the doped two--dimensional Hubbard
model, and here we will examine why this happens.

\begin{figure}
\centerline{\epsfysize=6.5cm \epsffile[-10 184 564 598] {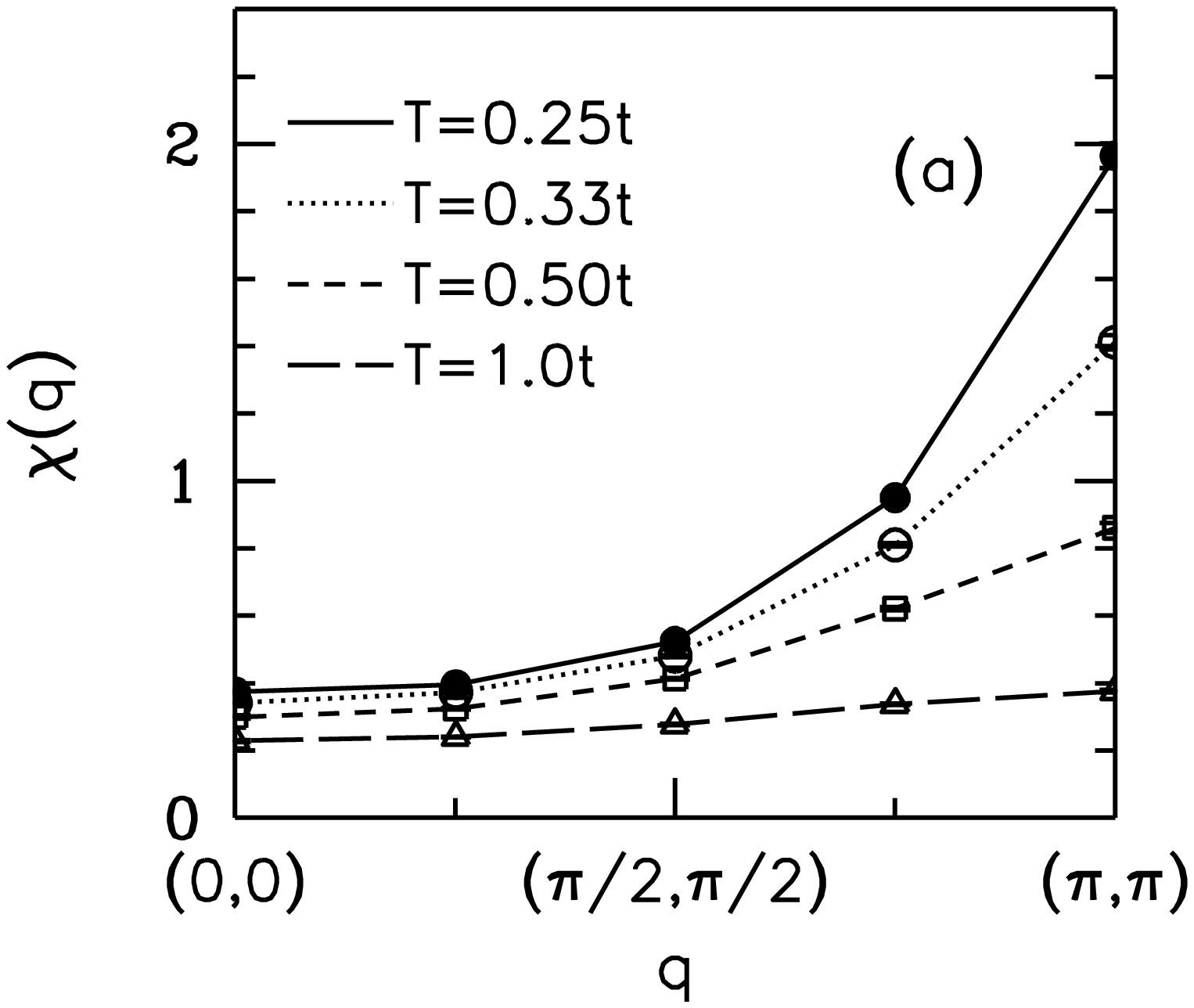}
\epsfysize=6.5cm \epsffile[78 184 652 598] {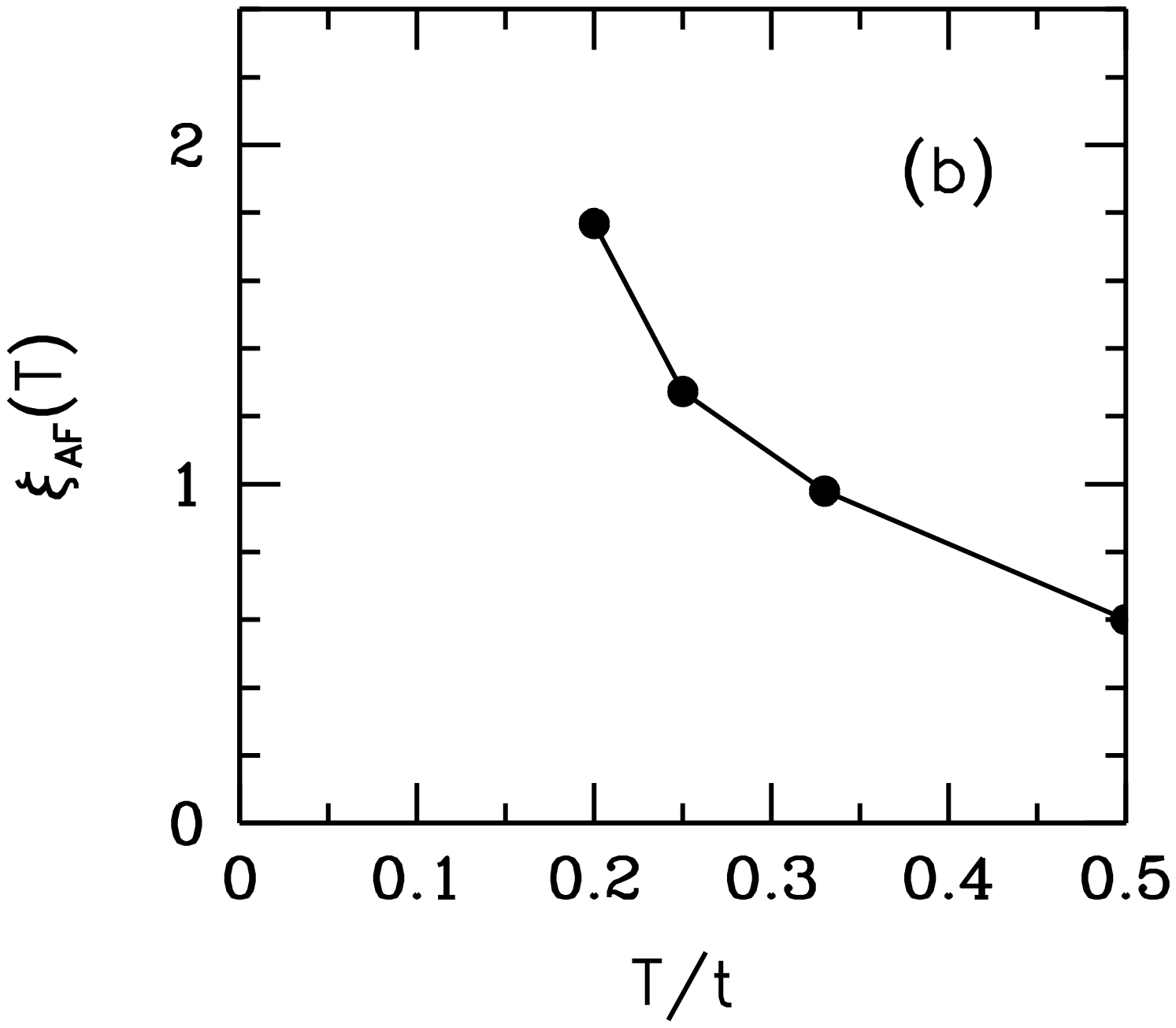}}
\caption{
(a) Magnetic susceptibility $\chi({\bf q})$ versus
{\bf q} along the (1,1) direction for various temperatures.
(b) Temperature dependence of the AF correlation length 
$\xi_{\rm AF}$ in units of the lattice spacing $a$.
These results are for an $8\times8$ lattice with $U/t=4$ and a filling
$\nang=0.875$.
\label{fig:chiq}}
\end{figure}

At half--filling, $\langle n_{i\uparrow} + n_{i\downarrow}\rangle = 1$,
the 2D Hubbard model develops long--range antiferromagnetic order as the
temperature goes to zero.
In the doped case, strong short--range AF correlations
develop as the temperature decreases below $J$.
This is clearly seen in the temperature dependence of the wave vector
dependent magnetic susceptibility
\begin{equation}
\chi({\bf q}) = {1\over N} \sum_{\lb}
  \int^\beta_0 d\tau\, 
      \left\langle m^-_{i+\ell}(\tau) m^+_i(0) \right\rangle 
  e^{-i{\bf q}\cdot \lb}. 
\label{eq:chiq}
\end{equation}
Here $m^+_i = c^{\dagger}_{i\uparrow} c_{i\downarrow}$ and
$m^-_{i+\ell}(\tau) = e^{H\tau} m^-_{i+\ell} e^{-H\tau}$,
where $m^-_{i+\ell}$ is the hermitian conjugate of 
$m^+_{i+\ell}$.
Monte Carlo results for $\chi({\bf q})$ versus $\q$ along the $(1,1)$
axis for an $8\times 8$ lattice with $U/t=4$ and a filling $\nang=0.875$
are shown in Fig.~\ref{fig:chiq}(a).
As the temperature decreases below $J=4\,t^2/U$, significant short--range 
dynamic AF correlations evolve.
Fig. \ref{fig:chiq}(b) shows the AF correlation length 
$\xi_{\rm AF}$ versus $T$.
Here, $\xi_{\rm AF}^{-1}$ is defined as the half--width at
half--maximum of $\chi(\q)$.

\begin{figure}
\centerline{\epsfysize=6.5cm \epsffile[-10 184 564 598] {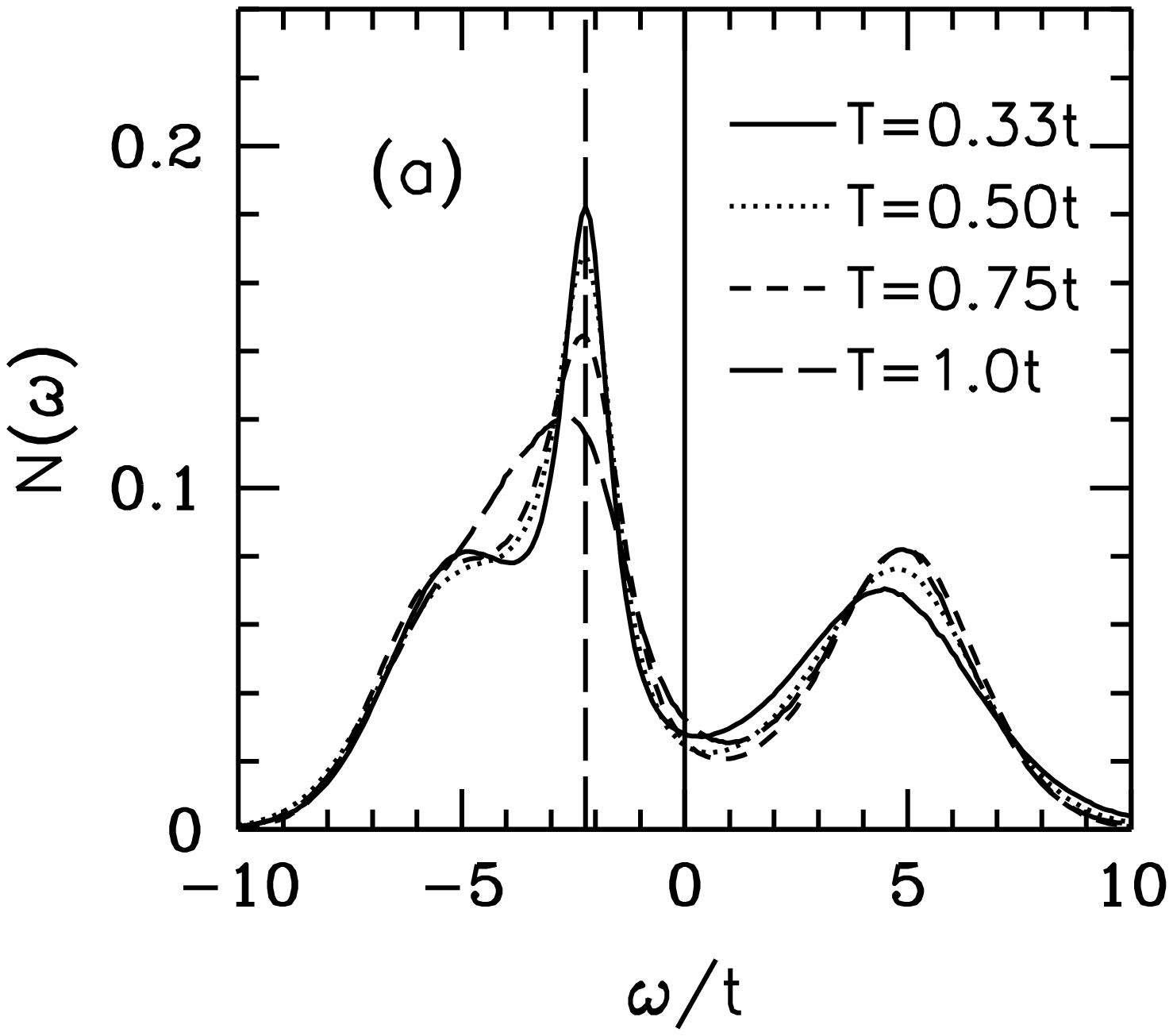}
\epsfysize=6.5cm \epsffile[78 184 652 598] {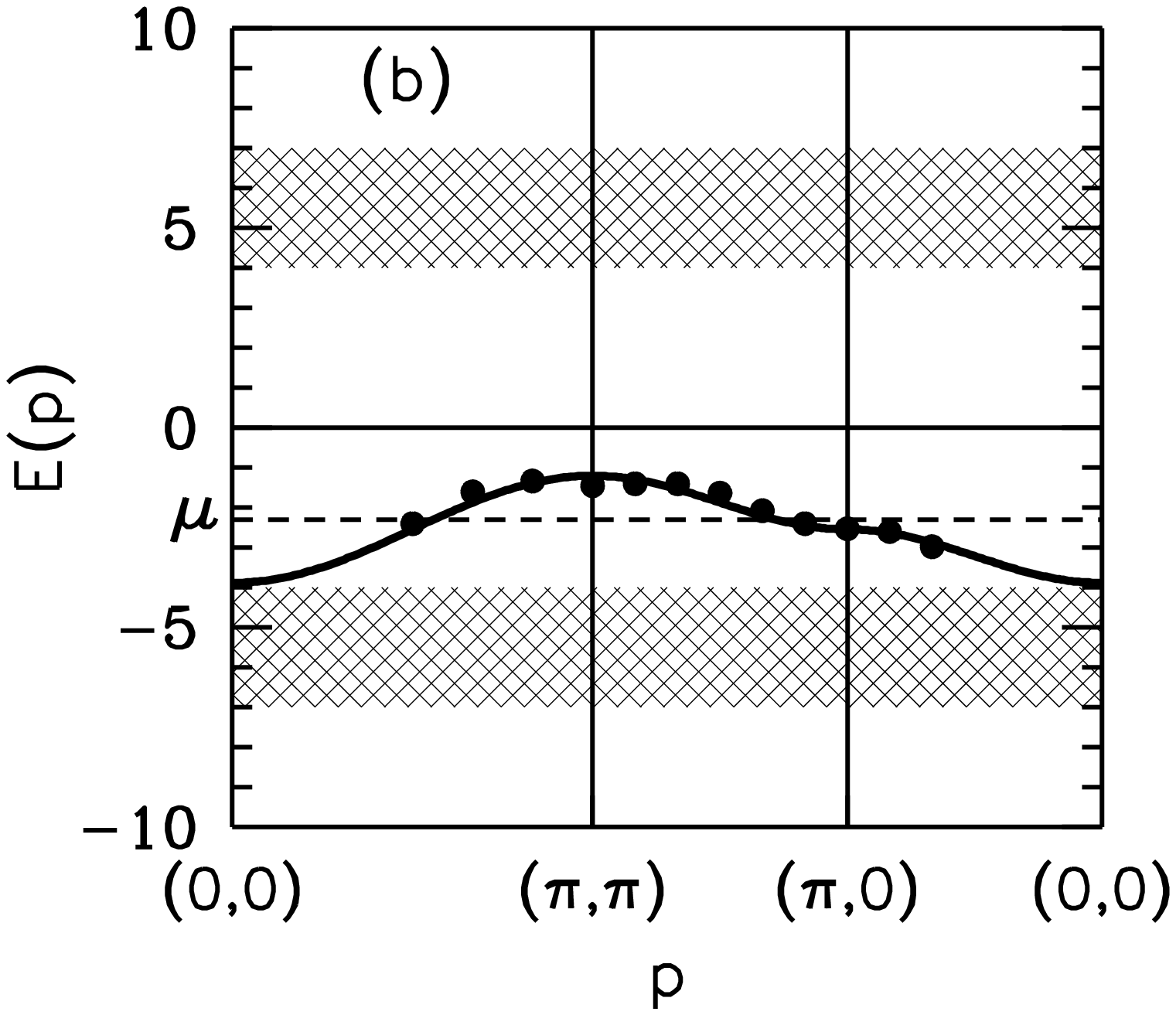}}
\caption{
(a) Evolution of the single--particle density of states with
temperature for $U/t=8$ and $\nang=0.875$.
(b) Dispersion of the quasiparticle peak in the spectral
weight versus {\bf p}.
The solid points mark the low--energy peaks of $A(\p,\omega)$ shown in
Fig.~3, 
and the solid curve represents an estimate of the quasiparticle
dispersion using these data and Lanczos results for {\bf p} near (0,0).
The broad darkened areas represent the incoherent spectral weight in the
upper and lower Hubbard bands.
The horizontal dashed line denotes the chemical potential $\mu$.
\label{fig:Nw}}
\end{figure}

As these AF correlations develop, the single-particle
spectral weight~\cite{Bul2}
\begin{equation}
A({\bf p},\omega) = -{1\over\pi}\,\im\,G({\bf p},
i\omega_n\to\omega+i\delta) 
\label{eq:Apw}
\end{equation}
and the density of states
$N(\omega) = {1\over N} \sum_{\p} A({\bf p},\omega)$
also change.
Figure \ref{fig:Nw}(a) shows $N(\omega)$ for $U/t=8$ and $\nang=0.875$.
As the temperature is lowered, a peak appears on the upper edge of the
lower Hubbard band.

\begin{figure}
\centerline{\epsfysize=7.5cm \epsffile[-207 144 367 718] {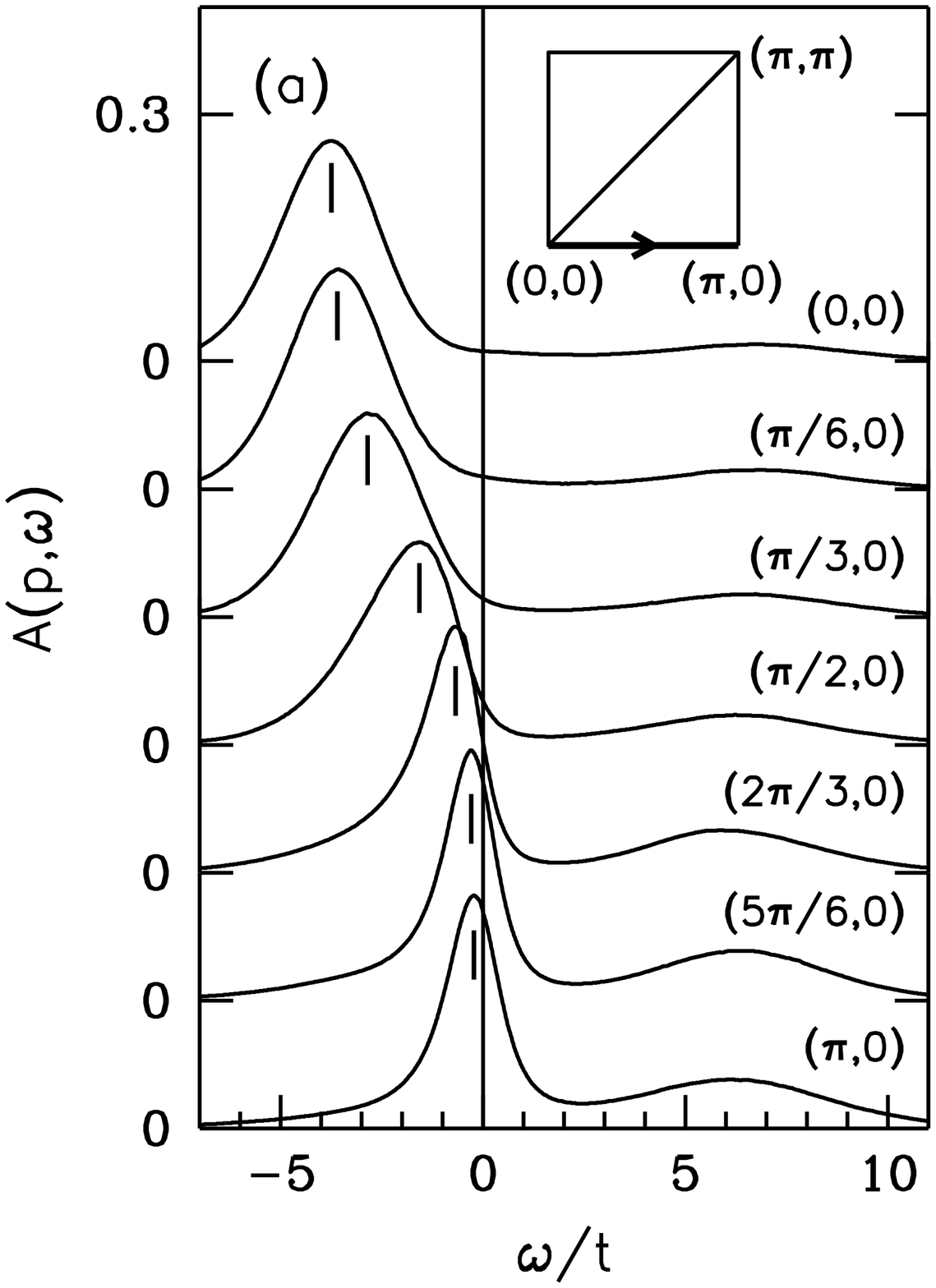}
\epsfysize=7.5cm \epsffile[18 144 592 718] {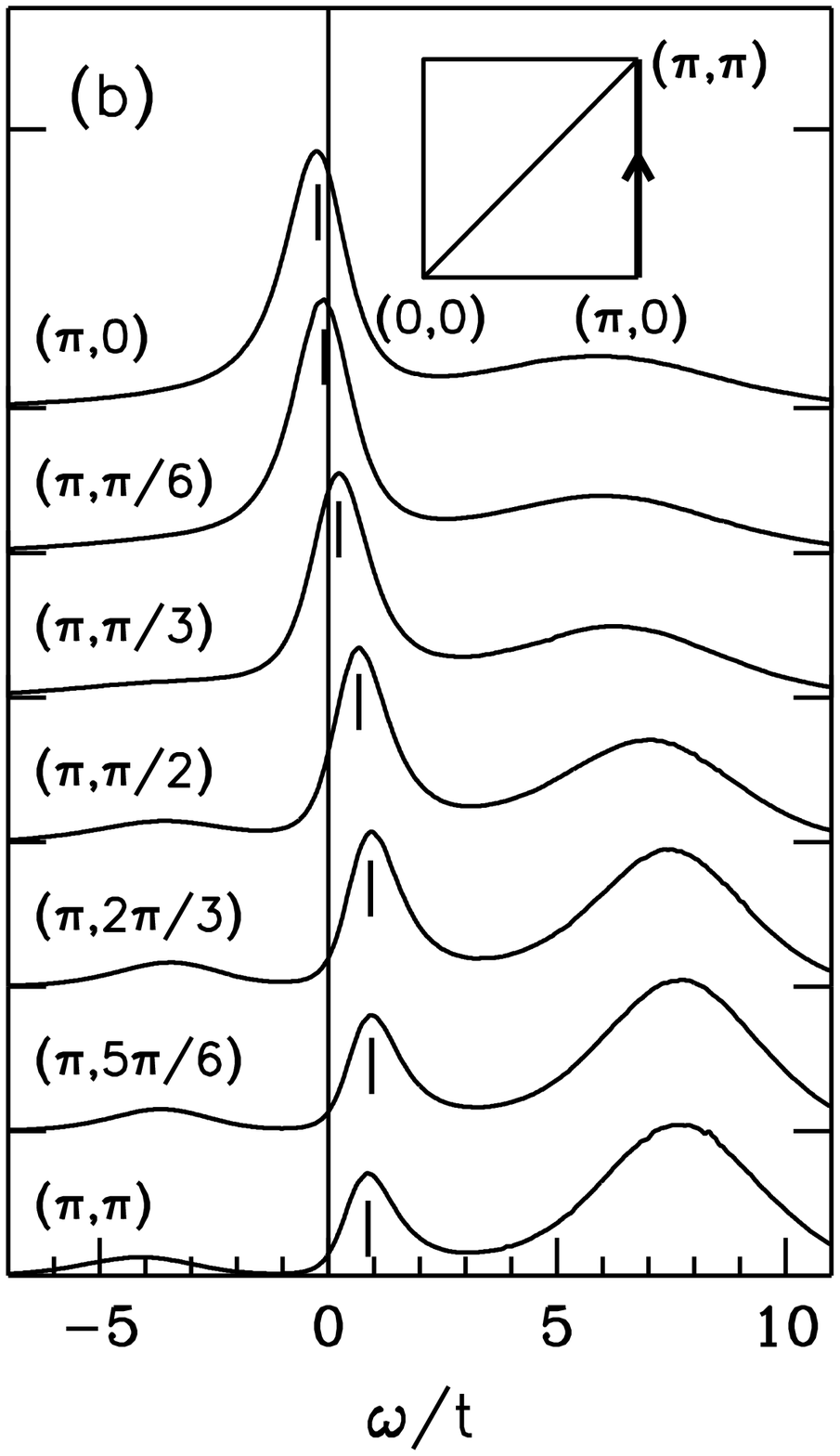}
\epsfysize=7.5cm \epsffile[243 144 817 718] {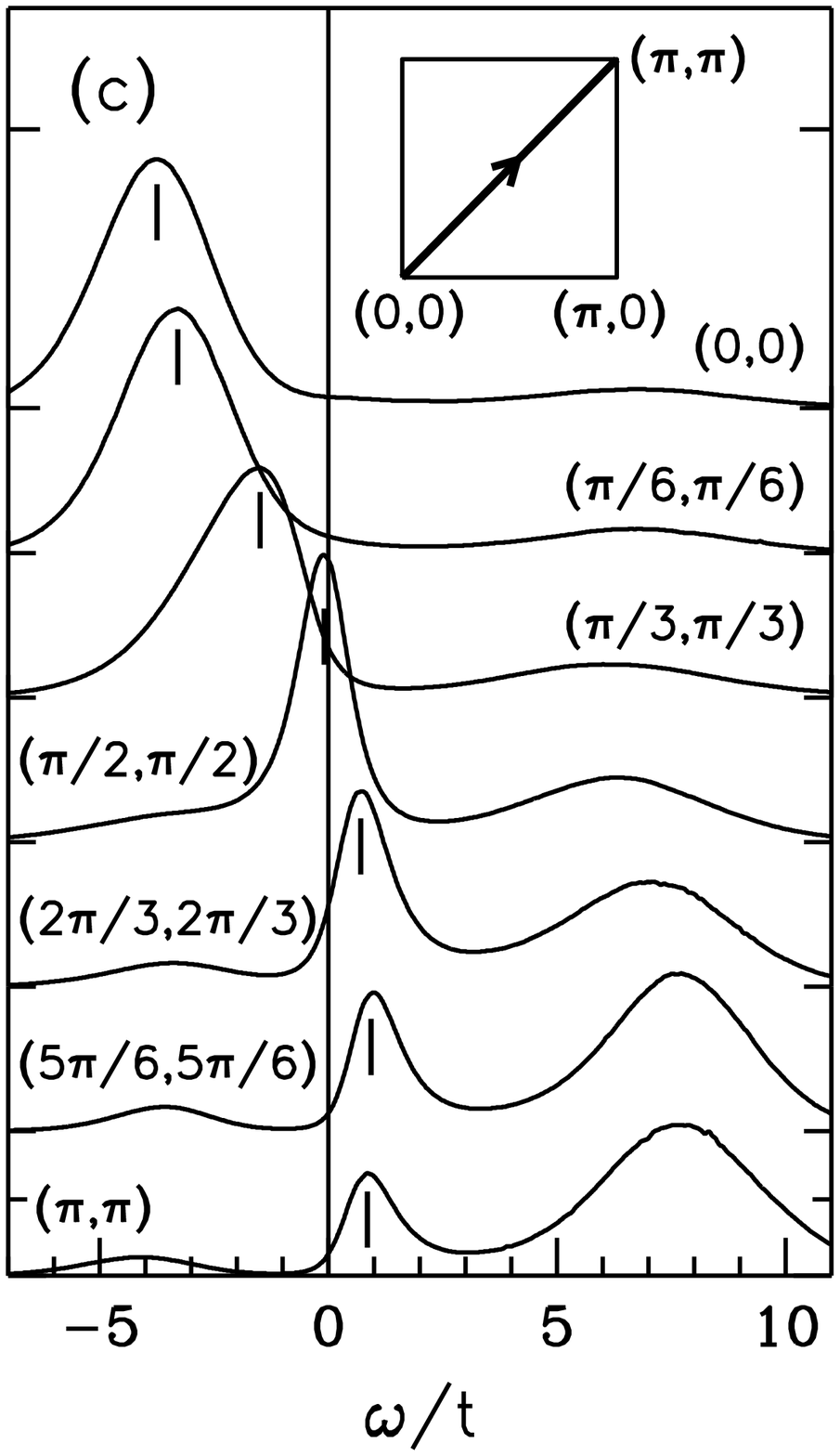}}
\caption{
Single--particle spectral weight along various cuts 
in the Brillouin zone is shown for $U/t=8$ and $\nang=0.875$ on a $12\times12$
lattice at $T=0.5\,t$.
\label{fig:Apw}}
\end{figure}

This peak arises from a narrow quasiparticle band shown in the
single--particle spectral weight $A({\bf p},\omega)$ of 
Fig.~\ref{fig:Apw} and
plotted as the solid curve in Fig.~\ref{fig:Nw}(b).
As the momentum {\bf p} goes towards the $\Gamma$ point $(0,0)$, we
believe that the quasiparticle peak is obscured by the lower Hubbard
band because of the resolution of the maximum entropy technique which we
have used.
Indeed, at the $\Gamma$ point a separate quasiparticle peak is found
from Lanczos exact diagonalization 
on a $4\times4$ lattice~\cite{Lanc}.
Note that this implies a large hole--like Fermi surface.
As clearly evident in the spectral weight shown in 
Figs.~\ref{fig:Apw}(a) and (b),
the quasiparticle 
dispersion is anomalously flat near the $(\pi,0)$ corner.
As discussed by various authors, this reflects the influence of the
AF correlations on the quasiparticle excitation energy.
It is clear that the peak structure in $N(\omega)$ also arises from the
short--range AF correlations and is a many--body effect
rather than simply a non--interacting band Van Hove singularity.

Monte Carlo calculations~\cite{Bul3} have also been used to determine 
the singlet irreducible particle--particle vertex 
$\Gamma_{\rm IS} (p',-p', p, -p)$ 
in the zero center--of--mass momentum and energy channel
which gives the effective pairing interaction.
Here $p=(\p,i\om_n)$.
In Fig.~4(a), $\Gamma_{\rm IS}(q=p-p')$ is plotted for {\bf q} along the
$(1,1)$ direction and $i\omega_n= i\omega_{n'} = i\pi T$, corresponding
to zero Matsubara energy transfer.
Comparing Figs.~1(a) and 4(a), one clearly sees that the 
structure of the interaction and $\chi(\q)$ are
similar, both reflecting the development of the AF
correlations as $T$ is reduced below $J$.

\begin{figure}
\centerline{\epsfysize=6.5cm \epsffile[-10 184 564 598] {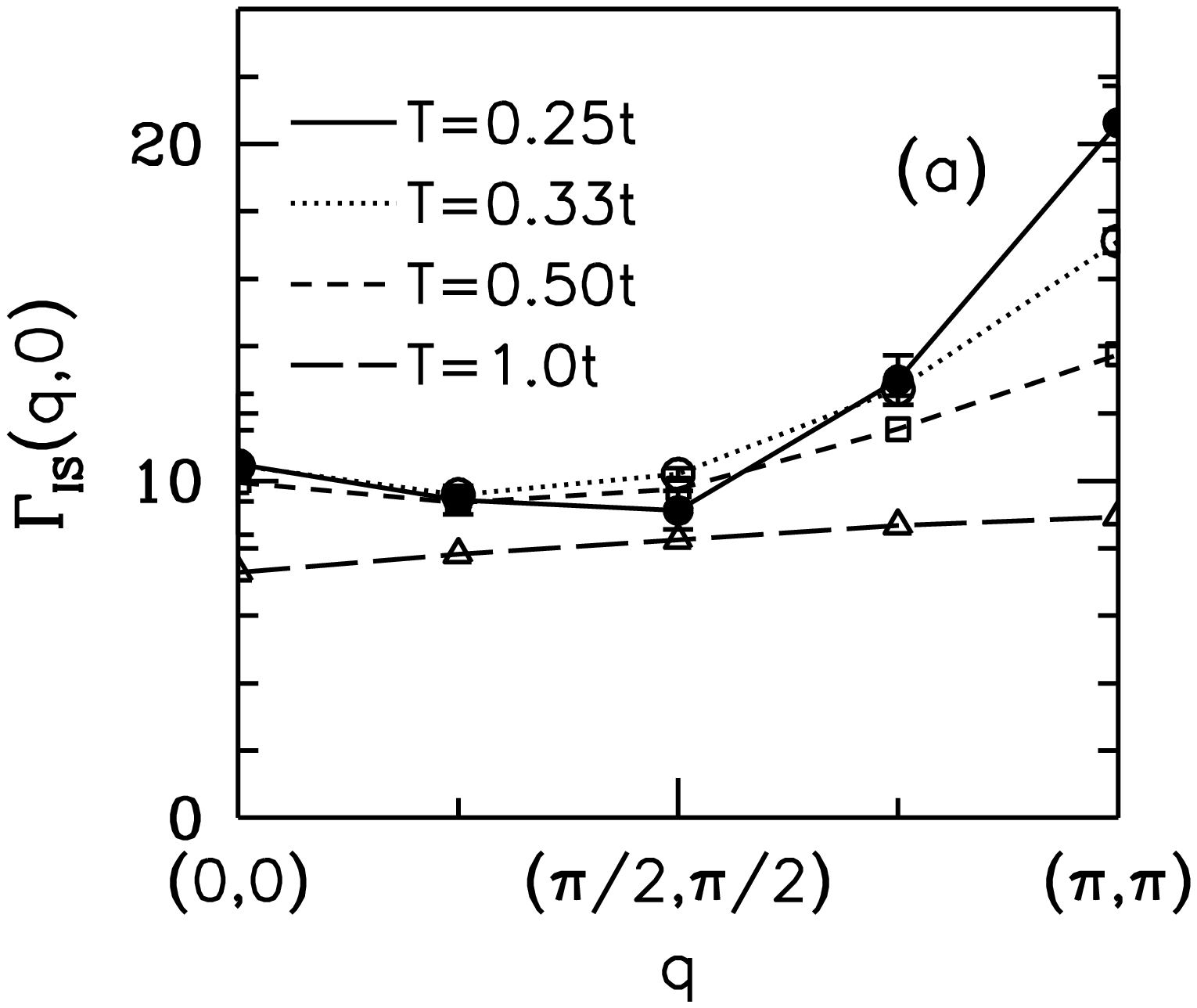}
\epsfysize=6.5cm \epsffile[78 184 652 598] {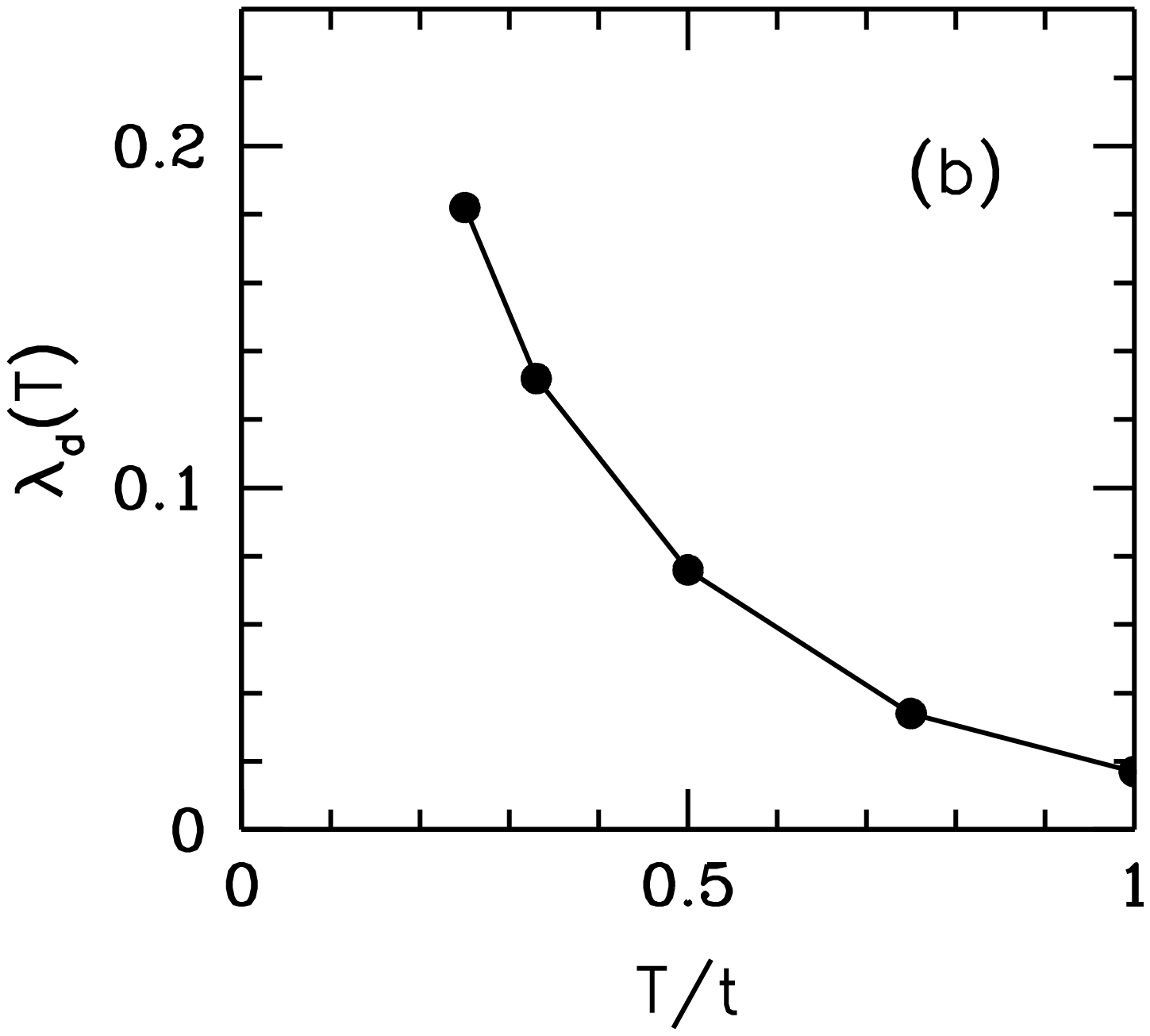}}
\caption{
(a) Singlet irreducible particle--particle vertex for zero
energy transfer $\Gamma_{\rm IS}({\bf q},
i\omega_m=0)$ versus {\bf q} along the (1,1) direction.
As the temperature decreases below $J=4\,t^2/U$, the strength of the
interaction is enhanced at large momentum transfer.
Note the similarity to $\chi({\bf q})$ in Fig.~1(a).
(b) Temperature dependence of the $\dxy$ eigenvalue of the 
Bethe--Salpeter equation.
These results are for $U/t=4$ and $\nang=0.875$.
\label{fig:gamI}}
\end{figure}

Given the Monte Carlo results for the irreducible
particle--particle vertex $\Gamma_{\rm IS}(p',-p',p,-p)$
in the zero energy and center--of--mass momentum channel,
and the single--particle Green's function $G(\p,i\om_n)$,
the Bethe--Salpeter equation for the particle--particle channel is 
\begin{equation}
\lambda_{\alpha}\phi_{\alpha}(p) = - {T\over N}
\sum_{p'} \Gamma_{\rm IS}(p,-p,p',-p') |G(p')|^2 \phi_{\alpha}(p').
\label{eq:bs}
\end{equation}
Here, the sum on $p'$ is over both $\p'$ and $\om_{n'}$.
In the parameter regime that the Monte Carlo simulations
are carried out, the leading singlet eigenvalue is in the 
$\dxy$ channel.  Fig. 4(b) shows the temperature 
dependence of the $\dxy$ eigenvalue.


The development of both the low--energy quasiparticle dispersion and the
peak in the singlet particle--particle vertex at large momentum transfers
arises from the growth of short--range AF correlations as
$T$ decreases below $J$.
As is known~\cite{289}, for the large Fermi surface associated with the 
observed quasiparticle dispersion, a particle--particle vertex which 
{\it increases\/} at large momentum transfer favors $\dxy$ pairing.
Note that the tendency for $\dxy$ pairing does not require a particularly 
sharp or narrow peak in
$\Gamma_{\rm IS}({\bf q})$ for ${\bf q} = \pp$, but rather simply
sufficient weight at large momentum transfers.
Thus it is the strong short--range AF correlations which
lead to the formation of $\dxy$ pairing correlations in the Hubbard
model.

\section*{Acknowledgments}
The work reviewed here was carried out with N.~Bulut and S.R.~White.
I would also like to acknowledge many useful discussions with
J.R.~Schrieffer.
This work was supported in part by the National Science Foundation under
grant No.~DMR92--25027.
The numerical computations reported in this paper
were carried out at the San Diego Supercomputer Center.



\section*{References}
\begin{thebibliography}{99}

\bibitem{Sch} J.R.~Schrieffer, 
\Journal{\em Solid State Commun.}{92}{129}{1995};
D.J.~Scalapino, \Journal{\em Phys.\ Reports}{250}{329}{1995};
D.~Van Harlingen, \Journal{\em Rev.\ Mod.\ Phys.}{67}{515}{1995}.

\bibitem{Dag} E.~Dagotto, A.~Moreo, R.L.~Sugar, and D.~Toussaint, 
\Journal{\PRB}{41}{811}{1990}.

\bibitem{Par} A.~Parola, S.~Sorella, M.~Parrinello, and E.~Tosatti, 
\Journal{\PRB}{43}{6190}{1991}.

\bibitem{Noack} R.M.~Noack, S.R.~White, and D.J.~Scalapino, 
\Journal{\PRL}{73}{882}{1994};
R.M.~Noack, S.R.~White, and D.J.~Scalapino, 
\Journal{\em Europhys.\ Lett.}{30}{163}{1995}.

\bibitem{Bul2} N.~Bulut, D.J.~Scalapino, and S.R.~White,
\Journal{\PRB}{50}{7215}{1994}.

\bibitem{Lanc} E. Dagotto, A. Moreo, F. Ortolani, J. Riera, and
D.J. Scalapino,
\Journal{\PRL}{67}{1918}{1991};
E. Dagotto, F. Ortolani, and D.J. Scalapino,
\Journal{\PRB}{46}{3183}{1992}.

\bibitem{Bul3} N.~Bulut, D.J.~Scalapino, and S.R.~White,
\Journal{\PRB}{50}{9623}{1994}.

\bibitem{289} D.J.\ Scalapino, 
\Journal{{\em Physica} C}{235--240}{107}{1994}.

\end{thebibliography}
\end{document}